\documentclass[10pt]{article}

\usepackage{amsmath,amssymb,amsthm}
\usepackage{amsxtra}
\usepackage{amsfonts}
\usepackage{amssymb}
\usepackage{url}
\usepackage{hyperref}
\usepackage{esint}
\usepackage{eucal}
\usepackage{mathrsfs}
\usepackage{cite}
\usepackage{graphicx,graphics}
\newcommand\fig[1] {{\rm Figure}~\ref{fig:#1}}
\newcommand\labfig[1] {\label{fig:#1}}

\newcommand\labsect[1] {\label{sect:#1}}

\newcommand\eq[1] {(\ref{#1})}
\newcommand{\bfm}[1]{\mbox{\boldmath ${#1}$}}

\newcommand{\nonum}{\nonumber \\}
\newcommand{\beqa}{\begin{eqnarray}}
\newcommand{\eeqa}[1]{\label{#1}\end{eqnarray}}
\newcommand{\beq}{\begin{equation}}
\newcommand{\eeq}[1]{\label{#1}\end{equation}}
\newcommand{\Grad}{\nabla}
\newcommand{\Div}{\nabla \cdot}

\newcommand{\Real}{\mathop{\rm Re}\nolimits}
\newcommand{\Imag}{\mathop{\rm Im}\nolimits}

\newcommand{\Md}{\partial}

\newcommand{\Ga}{\alpha}
\newcommand{\Gb}{\beta}

\newcommand{\Ge}{\epsilon}
\newcommand{\Gve}{\varepsilon}
\newcommand{\Gf}{\phi}

\newcommand{\Gk}{\kappa}

\newcommand{\Gn}{\eta}
\newcommand{\Gm}{\mu}

\newcommand{\Gt}{\theta}
\newcommand{\Gvt}{\vartheta}
\newcommand{\Gr}{\rho}

\newcommand{\Gj}{\tau}

\newcommand{\Go}{\omega}

\newcommand{\GL}{\Lambda}

\newcommand{\GO}{\Omega}

\newcommand{\BGve}{\bfm\varepsilon}

\newcommand{\BGn}{\bfm\eta}
\newcommand{\BGm}{\bfm\mu}
\newcommand{\BGv}{\bfm\nu}

\newcommand{\BGr}{\bfm\rho}

\newcommand{\BGs}{\bfm\sigma}

\newcommand{\BGG}{\bfm\Gamma}
\newcommand{\BGL}{\bfm\Lambda}

\newcommand{\BGY}{\bfm\Psi}

\newcommand{\CE}{{\cal E}}

\newcommand{\CJ}{{\cal J}}

\newcommand{\CT}{{\cal T}}

\newcommand{\BCC}{{\bfm{\cal C}}}
\newcommand{\BCD}{{\bfm{\cal D}}}

\newcommand{\BCS}{{\bfm{\cal S}}}

\newcommand{\BCV}{{\bfm{\cal V}}}

\newcommand{\bpm}{\begin{pmatrix}}
\newcommand{\epm}{\end{pmatrix}}

\def\Ba{{\bf a}}
\def\Bb{{\bf b}}

\def\Be{{\bf e}}
\def\Bf{{\bf f}}

\def\Bh{{\bf h}}

\def\Bj{{\bf j}}
\def\Bk{{\bf k}}

\def\Bm{{\bf m}}

\def\Bp{{\bf p}}
\def\Bq{{\bf q}}
\def\Br{{\bf r}}
\def\Bs{{\bf s}}

\def\Bu{{\bf u}}
\def\Bv{{\bf v}}

\def\Bx{{\bf x}}
\def\By{{\bf y}}

\def\BA{{\bf A}}
\def\BB{{\bf B}}

\def\BD{{\bf D}}
\def\BE{{\bf E}}
\def\BF{{\bf F}}
\def\BG{{\bf G}}

\def\BI{{\bf I}}
\def\BJ{{\bf J}}
\def\BK{{\bf K}}
\def\BL{{\bf L}}
\def\BM{{\bf M}}

\def\BP{{\bf P}}
\def\BQ{{\bf Q}}

\def\BT{{\bf T}}
\def\BU{{\bf U}}

\def\BZ{{\bf Z}}

\usepackage{graphics}
\usepackage{amssymb}

\title{A unifying perspective on linear continuum equations prevalent in science. Part II: Canonical forms for time harmonic equations}
\author{}
\date{}
\begin{document}
\maketitle
\vskip -.5cm
\centerline{\large Graeme W. Milton}
\centerline{Department of Mathematics, University of Utah, USA -- milton@math.utah.edu.}
\vskip 1.cm
\begin{abstract}
  Following some past advances, 
  we reformulate a large class of linear science equations in the format
  of the extended abstract theory of composites so that we can apply this theory to
  better understand and efficiently solve those equations.  Here in part  II  we elucidate the
  form for many time harmonic equations that do not involve higher order gradients. 
\end{abstract}
\section{Introduction}
\setcounter{equation}{0}
\labsect{1}
As in Part I \cite{Milton:2020:UPLI}, we are interested in giving examples of linear science equations that can be expressed in the form
\beq \BJ(\Bx)=\BL(\Bx)\BE(\Bx)-\Bs(\Bx),\quad \BGG_1\BE=\BE,\quad\BGG_1\BJ=0,
\eeq{ad1}
as encountered in the extended abstract theory of composites, where $\Bx=(x_1,x_2,x_3)$ is the spatial variable, $\Bs(\Bx)$ is the source term, 
$\BL(\Bx)$ governs the material response, and the selfadjoint projection operator $\BGG_1$ acts locally in Fourier space:
if $\BGG_1$ acts on a field $\BF$ to produce a field $\BG$ then we have that $\widehat{\BG}(\Bk)=\BGG_1(\Bk)\widehat{\BF}(\Bk)$ in
which $\widehat{\BG}(\Bk)$ and $\widehat{\BF}(\Bk)$ are the Fourier components of $\BG$ and $\BF$, and $\Bk=(k_1,k_2,k_3)$ represents a point
in Fourier space. Until Part  VII \cite{Milton:2020:UPLVII} we consider these equations in a medium of infinite extent,
possibly, though not necessarily, periodic.

This Part II is largely based on the book \cite{Milton:2016:ETC} and the paper \cite{Milton:2009:MVP}, focused on expressing time harmonic linear
continuum equations of science
in the desired form. We emphasize that the results apply to more general time dependencies if we resolve our source into its temporal Fourier
or Laplace components, and then integrate the response to each component.

The fields in \eq{ad1} are square integrable over all space, or if periodic, integrable over the unit cell of periodicity. As in Part I, given any
two fields $\BP_1(\Bx)$ and $\BP_2(\Bx)$ in this space of fields, we define the inner product of them
to be
\beq (\BP_1,\BP_2)=\int_{\mathbb{R}^3}(\BP_1(\Bx),\BP_2(\Bx))_{\CT}\,d\Bx,
\eeq{innp}
where $(\cdot,\cdot)_{\CT}$ is a suitable inner product on the space $\CT$
such that the projection $\BGG_1$ is selfadjoint with
respect to this inner product, and thus the space $\CE$ onto which
$\BGG_1$ projects is orthogonal to the space $\CJ$ onto which
$\BGG_2=\BI-\BGG_1$ projects. When we have periodic fields in periodic media the integral in \eq{innp} should be taken over the unit cell of periodicity.
We allow for nonperiodic fields
in periodic media provided they are square integrable over all space. At each point $\Bx=(x_1,x_2,x_3)$ the fields take values in a space $\CT$ of supertensors,
by which we mean a finite collection of scalars, vectors, and tensors.

To recall from Part I, as noted for example by Strang \cite{Strang:1986:IAMB}, a large class of  science equations can be expressed in the form
\beq  \BD(\Grad)^\dagger\BL\BD(\Grad)\BGY=\Bf, \eeq{conve1}
where $\BGY$ and $\Bf$ are the potentials and source, $\BD(\Grad)$ is a differential operator and $\BD(\Grad)^\dagger$
its adjoint. Setting
\beq \BGG_1(\Bk)={\BD}(i\Bk)[{\BD}(i\Bk)^\dagger{\BD}(i\Bk)]^{-1}{\BD}(i\Bk)^\dagger, \quad \Bf=\BD(\Grad)^\dagger\Bs,\quad \BE=\BD(\Grad)\BGY,
\quad \BJ=\BL\BE-\Bs,
\eeq{conve2}
one sees how this broad class of equations can be expressed in the form \eq{ad1}. We will find that for many wave equations $\BD^\dagger\BL\BD$
takes the form
\beq \BD(\Grad)^\dagger\BL\BD(\Grad)=\underline{z}\BI-\BD(\Grad)^\dagger\underline{\BB}\BD(\Grad),
\eeq{conve3}
where the parameter $\underline{z}$ may be $\Go^2$, where $\Go$ is the frequency or the energy $E$ for quantum waves. Then \eq{conve1}
has a solution in terms of the associated resolvent:
\beq \BGY=[\underline{z}\BI-\BD(\Grad)^\dagger\underline{\BB}\BD(\Grad)]^{-1}\Bf.
\eeq{conve4}

We will come across examples where $\BL$ has a nontrivial null space.
If possible, as discussed in Part I,  one may be able to shift $\BL(\Bx)$ by a
multiple $c$ of a ``null-$\BT$ operator''
$\BT_{nl}(\Bx)$ (acting locally in real space), defined to have the
property that $\BGG_1\BT_{nl}\BGG_1=0$ (implying that the
associated quadratic form (possibly zero) is a ``null-Lagrangian'').
 and then if there are infinities
in the resultant $\BL(\Bx)$ one may be able to shift  $\BL^{-1}(\Bx)$
by a multiple of a ``null-$\BT$ operator'' $\widetilde{\BT}_{nl}(\Bx)$
satisfying
$\BGG_2\widetilde{\BT}_{nl}\BGG_2=0$ to remove its degeneracy. 

Some of the equations we discuss have a generalized form that is particularly associated with metamaterials. These often have resonating subelements
leading to unusual macroscopic behavior that these equations capture. Capriz foreshadowed these developments when he introduced the idea of materials
with latent microstructure back in 1985 \cite{Capriz:1985:CLM}. The multifield model of Capriz embodies at each point $(\Bx,t)$ in
space time a coarse grained morphological descriptor field $\BGv(\Bx,t)$ that may not necessarily be observable, but which interacts with
observable fields in a dynamical systems way. Thus the observable fields appear to be governed by equations that are nonlocal in time, and
possibly nonlocal in space too. This viewpoint is a natural generalization of the abundant ``hidden variable'' theories and theories with
abstract order parameters \cite{Landau:1980:SP}, such as those used in the theory of 
liquid crystals and superfluids \cite{Chandrasekhar:1992:LC, Mermin:1979:TTD}. Their history and
subsequent developments are summarized and explored further in the paper of Mariano and Stazi \cite{Mariano:2005:CAM}.

There is an avalanche of papers, and associated
reviews on the subject of metamaterials. We just mention the books
\cite{Engheta:2006:EMP, Cai:2010:OM,  Banerjee:2011:IMW, Craster:2013:AMN, Pai:2015:TDA, DelIsola:2020:DCM, Lakes:2020:CM},
the articles \cite{Eleftheriades:2009:ETL, McPhedran:2011:MM,
  Milton:2016:AM, Kadic:2019:M},
and references therein. Many definitions of metamaterials have been proposed, most departing significantly from Walser's original
definition \cite{Walser:2001:EM}. We use the definition \cite{Kadic:2019:M}:
\vskip 5mm

{\it Metamaterials are rationally designed composites made of tailored building blocks that are composed of one or more constituent bulk materials.
  The metamaterial properties go beyond those of the ingredient materials, qualitatively or quantitatively. The properties of metamaterials can be
  mapped onto effective medium parameters.}
\vskip 5mm

 A beautiful example 
are the fantastic metamaterials made from parallelogram mechanisms using razor blades  constructed by
the architect/artist Boris Stuchebryukov (see the  videos
\url{https://www.youtube.com/watch?v=l8wrT2YB5s8} and \url{https://www.youtube.com/watch?v=AkUn8nFd8mk}).
i
The subject of metamaterials has a long history with contributions from many people. For example:
\begin{itemize}
\item the Romans knew that small metal particles embedded in glass could cause it to have unexpected colors (as in the 4th century
Lycurgus cup displayed at the British museum)
explained later in 1904 on the basis of optical resonances of the particles \cite{MaxwellGarnett:1904:CMG} due to the negative electrical permittivity
of metals.
Further experiments
demonstrate that hollow or ellipsoidal particles give a tunable range of colors \cite{Lai:2007:NOS} as predicted by the well known formulas
for the polarizability of hollow spheres and ellipsoids.

\item Faraday in 1837 recognized that arrays
of conducting inclusions could give an artificial dielectric response, a model he thought may account for the dielectric properties of all
materials \cite{Landauer:1978:ECI};

\item Bose in 1898 recognized that metamaterials with twisted jute could rotate the plane of polarization of electromagnetic waves \cite{Bose:1898:RPP},
  in the same way that  molecules such as sugar cause such rotations (known as optical activity).
Since then, a large variety of "bianisotropic" metamaterials have been discovered that exhibit unusual electromagnetic 
properties \cite{Serdyukov:2001:EBAM} including, in 1948, Tellegen's isotropic nonreciprocal magnetoelectric metamaterial \cite{Tellegen:1948:GNE};

\item At least since the second world war it was known that bubbles in water can form a metamaterial screen for absorbing sound, in particular the noise from submarine
  propellers \cite{Loye:1948:SAB} -- try ringing a glass without water, with water, and then with alka seltzer added to the water to make bubbles to see
  its effectiveness;

\item that split ring resonators could give rise to artificial magnetism was known to Schelkunoff and Friis in 1952 \cite{Schelkunoff:1952:ATP};

\item going beyond Faraday, Brown in 1953 \cite{Brown:1953:ADR}
  found that arrays of conducting rods or arrays of conducting plates with holes could give rise to artificial dielectrics with a refractive
  index less than one.
  Exploring this further Rotman in 1962 \cite{Rotman:1962:PSA}
  found the response for perfectly conducting
  rods was exactly the same as
  for a lossless plasma (which clearly has a negative electrical permittivity
  below the plasma frequency);

\item that new terms could enter the homogenized equations in high contrast media was known to Barenblatt, Zheltov, and Kochina in 1960
  \cite{Barenblatt:1960:FEF, Barenblatt:1960:BCT}.
  This type of homogenization, called double porosity scaling provides the mathematical basis for understanding the behavior of many
  metamaterials: see the excellent paper \cite{Smyshlyaev:2009:PLE} and references therein;
  
\item  that effective mass densities (by which we mean for small time harmonic vibrations)
  need not be a volume average of the local density was known to Berryman in 1980 \cite{Berryman:1980:LWPa};

\item that effective mass densities could be anisotropic was known to Schoenberg and Sen in 1983 \cite{Schoenberg:1983:PPS};

\item that effective mass densities could be negative (and anisotropic and frequency dependent) was known to
  Aurialt and Bonnet in 1985 \cite{Auriault:1985:DCE};

\item that one can get materials that behave neither like dielectrics nor conductors at low frequencies was known
to experimentalists measuring the electrical permittivity of porous rocks containing electrically conducting salt water,
who found that the {\it real part} of the complex effective electrical permittivity can diverge as the frequency tends to zero ---
this can be accounted for by a continuum of resonances of the effective electrical permittivity as a function of frequency \cite{Stroud:1986:AMD}
(and, in a similar vein, the viscoelastic creep of cast Indian-Tin can stretch over 7 decades of frequency! \cite{Lakes:1996:VBI});

\item that composites, unlike most materials,  could have a negative Poisson's ration was discovered in experiments by Lakes in 1987 \cite{Lakes:1987:FSN}
  and rigorously proved in \cite{Milton:1992:CMP};
  
\item that, as discovered in 1993 \cite{Nicorovici:1993:TPT}, metamaterial
  arrays of cylinders
  with a core material coated with a material
  having a permittivity $\Gve/\Gve_0=-1$, surrounded by
  void with permittivity $\Gve_0$ have an
  equivalent quasistatic response as an array of cylinders of the core
  material having much larger radius.
  This led to the discovery of anomalous resonance \cite{Nicorovici:1994:ODP},
as reviewed in \cite{McPhedran:2020:RAR}, 
  where a source excites a resonant field with
  resonance confined to region dependent on the position of the source.
  Anomalous resonance can also produce ``ghost'' sources that mimic real
  sources \cite{Nicorovici:1994:ODP, Milton:1994:URP, Milton:2005:PSQ}.  It is
  associated with essential singularities of the response function rather
  than with the poles associated with normal resonance
  \cite{Nicorovici:1993:TPT, Ammari:2013:STN}; 

 \item that one could combine materials all with positive thermal expansions to get a metamaterial having negative thermal expansion
  was discovered by in 1996 by Lakes \cite{Lakes:1996:CSS} and Sigmund and Torquato \cite{Sigmund:1996:CET}.
  The effect was then discovered in 1998 by Baughman, Stafstr{\"o}m, Cui, and Dantas for the analogous
  problem of poroelasticity \cite{Baughman:1998:MNC}, and experimentally verified \cite{Qu:2017:EMN}--- the equivalent of blowing into a balloon filled with a
  metamaterial and seeing the balloon contract;

\item that the effective magnetic permeability could be negative was shown in experiments by Lagarkov, Semenenko, Chistyaev, Ryabov, Tretyakov, and Simovski
  in 1997  \cite{Lagarkov:1997:RPB}.

\end{itemize}

Shelby, Smith and Schultz \cite{Shelby:2001:EVN} combined the split ring
metamaterial elements of Schelkunoff and Friis (that give negative permeability)
with the highly conducting rod elements of Brown (that give
negative permittivity) to obtain a material with 
negative permeability $\Gm$ and negative permittivity $\Gve$ that
they showed had a  negative effective refractive index.
They have the unusual property, studied as far back as 1904, 
  \cite{Lamb:1904:GV, Schuster:1904:ITO} that the phase  velocity
  (governing the movement of wave crests) is opposite to the group velocity (governing the movement of energy). Veselago had realized that materials with
  negative $\Ge$ and $\Gm$ would propagate waves and have a negative refractive
  index \cite{Veselago:1967:ESS}. Moreover, he realized that a slab having
  $\Gm/\Gm_0=\Gve/\Gve_0=-1$ should function as a lens. Sir John Pendry used
  questionable analysis to claim that this lens would function as a perfect
  lens with subwavelength resolution \cite{Pendry:2000:NRM}.
  This triggered a surge of interest in metamaterials. The essential mechanism for superlensing turned out to be anomalous resonance
  \cite{Nicorovici:1994:ODP, Milton:2005:PSQ}: the anomalous
  resonance sets the scale of resolution which points to its essential role
  (and this scale of resolution is not much beyond the Abbe diffraction limit
  for materials with realistic values of the imaginary parts of $\Ge$
  and $\Gm$).   The quasistatic approximation still holds true for
  the anomalously resonant
  fields of the slab lenses, because the field gradients are so huge.
  An interesting aspect of anomalous resonance
  is that it is associated
  with cloaking \cite{Milton:2006:CEA, Ammari:2013:STN, Nguyen:2017:CAO}.
  In fact even the claimed perfect lens with $\Gm/\Gm_0=\Gve/\Gve_0$
  exactly $-1$ at one frequency can act to cloak constant power
  sources, rather than
  perfectly image them: as time goes on increasing fractions of 
  the power from the source get funneled
  into the region of  anomalous resonance, with the source fading from
  view (both in front and behind the lens!)
  \cite{Milton:2006:OPL}. This dimming of the source
  is demanded by conservation of
  energy and points to the serious gaps in the original analysis that are
  glossed over in most books, reviews, and Wikipedia entries
  that continue to mislead readers.

What we now call metamaterials have been labeled by various terms in the past:  advanced materials, architectured materials, artificial materials,
complex materials, designer matter, generalized continua, multifield materials, multiscale materials, multiphysics materials, negative moduli materials, optimized materials, properties on demand, smart materials, advanced materials, as listed in \cite{DelIsola:2020:MW, Kadic:2019:M} along with
many references. This wealth of names is a testament to the enormous body of work on this subject, even back in the last millennium.

Metamaterials require one to combine materials with a high contrast in their properties --- otherwise the effective properties will be qualitatively
similar to those of the constituents. From an homogenization viewpoint there are two possibilities.
One can consider families of materials with $\BL(\Bx)$ taking the form
\beq \BL(\Bx)=\underline{\BL}_\Ge(\Bx,\Bx/\Ge), \eeq{hv}
generated from functions $\underline{\BL}_\Ge(\Bx,\By)$ that are periodic in the ``fast variable'' $\By$. The homogenized equations apply in the limit as
$\Ge\to 0$. One possibility is that $\underline{\BL}_\Ge(\Bx,\By)$ is independent of $\Ge$ but itself has high contrast. Then one gets interesting
effective moduli, but the homogenized equations take the same form
as the original equations. A second possibility, such
as associated with the double porosity scaling \cite{Barenblatt:1960:FEF, Barenblatt:1960:BCT, Smyshlyaev:2009:PLE} is that $\underline{\BL}_\Ge(\Bx,\By)$ depends on $\Ge$ with increasing contrast as $\Ge\to 0$.
Then terms that ordinarily give only second order contributions to the homogenization equations, and hence fade from significance as $\Ge\to 0$,
become first order, thus dramatically changing the homogenized behavior.
The homogenized equations can then take a different form from the
original equations.

The study of metamaterials should be distinguished from the study of the band
structure of materials, where the wavelength is of the order of the size of the period cell:
see, for example, the books \cite{Joannopoulos:2008:ISS, Deymier:2013:AMP, Phani:2017:DLM} and the article \cite{Gorishnyy:2005:SI}. 
However, we mention that homogenization theory carries over
to the vicinity of local minima, maxima, and saddle points of the dispersion
relation $\Go(\Bk)$, where it gives effective equations for the
modulation of waves and is known as high frequency homogenization.
This was observed for the Schr{\"o}dinger equation by
Bensoussan, Lions and Papanicolaou \cite{Bensoussan:1978:AAP} (see their
equations (4.33) and the discussion at the bottom of page 352)
and Birman and Suslina gave a rigorous treatment \cite{Birman:2006:HMP}
for the case of local minima or maxima of the dispersion relation
centered at $\Bk=0$. Craster, Kaplunov and Pichugin \cite{Craster:2010:HFH}
rediscovered the approach and coined the name high frequency homogenization.
They and their collaborators greatly extended the field and provided
supporting numerical calculations: see \cite{Harutyunyan:2016:HFH}
and references therein. 
Most interesting is the behaviour near saddle points where hyperbolic
effective equations describe the modulations, so that the radiation
concentrates along characteristic
lines \cite{Antonakakis:2014:HEP, Makwana:2015:WMM}.

We emphasize that, as we assume square integrability of the fields,
if the medium is not absorbing then the analysis only applies to bound
states. However, for numerical purposes or otherwise, one may introduce 
(at sufficiently large distances from the spatial region of
interest) perfectly matched layers or absorbing walls with graded material
properties to confine the fields. Later in section VII we will see
how radiation and scattering problems can be treated without having to
introduce these layers or walls.

As in Part I, the relevant physical fields are progressively defined, and we do not typically remind the reader of their
definitions in subsequent equations. Also,
to avoid taking unnecessary transposes, we let $\Div$ act on the first index of a field, and the action of $\Grad$ produces a field, the first index of
which is associated with $\Grad$.

\section{Time harmonic acoustics (Helmholtz equation)}
\setcounter{equation}{0}
\labsect{2}

The Helmholtz equation, as pertaining to acoustics, can be written in the form:
\beq \begin{pmatrix} P \\ \Grad P \end{pmatrix}=\BL\begin{pmatrix} i\Div\Bv \\ i\Bv\end{pmatrix}+\bpm 0 \\ \Bf\epm,
\eeq{15.11}
where $P$ is the complex pressure, and $\Bv$ the complex fluid velocity, and $\Bf$ is the complex volume force density such as due to an oscillatory electric
field or oscillatory electric field gradient if the fluid contains charged particles or electric dipoles. There could also be an oscillatory
source of pressure. This can be included by taking its gradient and adding it to $\Bf$. Oscillatory sources of fluid flow can be taken into
account too. 

We have
\beqa \BL=\begin{pmatrix}-\Gk/\Go & 0 \\ 0 & \Go\BGr\end{pmatrix}, \quad \BGG_1(\Bk)& = &\BZ(\Bk)\equiv \frac{{\BD(i\Bk)}{\BD(i\Bk)}^\dagger}{k^2+\Go^2}\quad\text{with  }\BD(\Grad)=\bpm \Grad \\ 1 \epm \nonum &=& 
\frac{1}{k^2+1}\begin{pmatrix}
\Bk\otimes\Bk & i\Bk \\
-i\Bk^T & 1
\end{pmatrix},
\eeqa{x10}
where now the effective mass density $\BGr(\Bx)$ and compressibility $\Gk(\Bx)$ may be frequency dependent and complex and $\BGr(\Bx)$ could be anisotropic.
Anisotropic and frequency dependent densities may sound strange, but they are a feature of metamaterials.
They are a consequence of the need to replace Newton's law $\BF=m\Ba$ with a law
where the force is convolved with the acceleration reflecting the fact that not all parts of the material respond immediately to an acceleration: there could
be some time lag \cite{Milton:2006:MNS}. In other words, not all the material moves in lockstep motion. 
This is directly analogous to the complex electrical permittivity being frequency dependent, resulting in part from the lag time it takes
(within a classical, not quantum, interpretation) for
electrons, atoms, or molecules to respond to an applied electric field due to their inertia. A formula for the effective mass density for fluids
containing solid inclusions that was not just a volume average of the solid and
fluid densities was proposed by Berryman as long ago as 1980 \cite{Berryman:1980:LWPa}. 
Its experimental confirmation was not until 2006 \cite{Mei:2006:EMD}.
In 1983 Schoenberg and Sen \cite{Schoenberg:1983:PPS} found anisotropic
layered fluids had an anisotropic effective density. In the context of the elastodynamic equations to be discussed shortly, Willis \cite{Willis:1985:NID}
introduced for composites an  effective density operator that was nonlocal (both in space and time), and Milton, Briane and Willis \cite{Milton:2006:CEM}
found simple models exhibiting a local (in space, not time) anisotropic density: see their Figure 3. Using a homogenization approach
Aurialt and Bonnet \cite{Auriault:1985:DCE, Auriault:1994:AHM} realized that the effective mass density could be frequency dependent, anisotropic,
or even negative. Negative effective masses were experimentally and theoretically confirmed in \protect{\cite{Sheng:2003:LRS, Liu:2005:AMP, Sheng:2007:DMD}}:
a simple model showing this is in  \fig{-1}.
A more rigorous study of Zhikov arrived at the same conclusion \cite{Zhikov:2004:GSE}. Smyshlyaev extended this analysis to elastic waves traveling
in extremely anisotropic media \cite{Smyshlyaev:2009:PLE}. 
Moreover, in considering lattices of gyroscopic spinners (a spinetic material)
Carta, Jones, Movchan, and Movchan \cite{Carta:2019:WPD} found the density matrix is not symmetric, but rather Hermitian (and in the presence of losses it would
not even be Hermitian).

In acoustic metamaterials the tensor $\BL(\Bx)$ can have some off-diagonal coupling \cite{Muhlestein:2016:MAH, Sieck:2017:OWC} 
directly analogous to the local Willis equations for elastodynamics, that will be discussed shortly. It is rather natural that such
coupling terms should also appear in acoustics since acoustics is a degenerate case of elastodynamics, with the shear modulus tending to zero.
Experimental evidence for such off-diagonal acoustic couplings has been obtained by Koo, Cho, Jeong, and Park \cite{Koo:2016:AOM} and by Muhlestein, Sieck,
Wilson, and Haberman \cite{Muhlestein:2017:EEW}. Similar off diagonal couplings are also a feature of the bianisotropic equations of
electromagnetism \cite{Serdyukov:2001:EBAM} where they first made
their appearance. If one is only interested in the pressure field $P(\Bx)$ then there is some ambiguity in writing the coupled equations. To see this,
suppose for simplicity that $\Bf=0$. If the coupled equations take the form
\beq \bpm \Bv \\ \Div\Bv \epm =\underbrace{\bpm \BA & \Ba \\ \Bb^T & c\epm}_{\BL'}\bpm \Grad P \\ P \epm, \eeq{amb}
then the pressure satisfies
\beq \Div\BA\Grad P=-\Div(\Ba P)+\Bb\cdot\Grad P +c P=(\Bb-\Ba)\cdot\Grad P +(c-\Div\Ba)P, \eeq{amb1}
implying that we are free to change or eliminate $\Ba(\Bx)$ and correspondingly adjust $\Bb(\Bx)$ and $c(\Bx)$
without changing the pressure field $P(\Bx)$. On the other hand, if we are also interested
in the velocity field $\Bv(\Bx)$, which is an observable quantity, then we do not have this freedom.

\begin{figure}[!ht]
\centering
\includegraphics[width=0.7\textwidth]{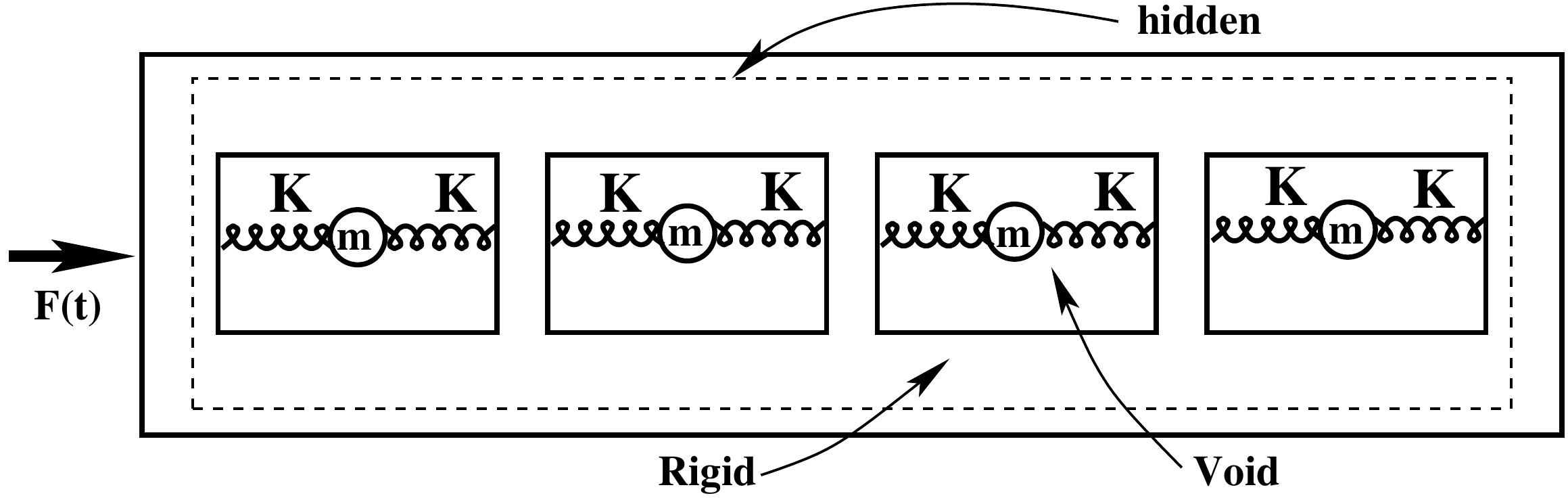}
\caption{A simple model illustrating the concept of negative effective mass, generalized to anisotropic effective mass. Inside the bar of mass $M_0$
  are $n$ hidden cavities each enclosing a mass $m$ attached to the sides of the cavity with springs having spring constant $K$. In pushing the bar with an oscillatory force
  $\BF(t)$ having frequency $\Go$ an elementary calculation shows the body responds like it was an object with mass $M=M_0+2Knm/(2K-m\Go^2)$. This
  effective mass is negative
just above the resonant frequency $\Go=\sqrt{2K/m}$. When this effective mass is negative the motion of the masses inside each cavity, and the associated momentum,
is opposite to that of the bar velocity, and the motion of the bar and the inertial forces are all that one sees from outside the bar.
If the springs have some damping then $K$ and $M$ become complex. One can similarly insert such rectangular cavities, containing masses joined
to the ends of the cavity by springs, into a rigid cube. If the cavities have both vertical orientations and horizontal orientations, with the
vertical springs having a different spring constant from the horizontal ones, the effective mass becomes anisotropic and can even have
both positive and negative eigenvalues. Adapted from Figure 1 in \protect{\cite{Milton:2007:MNS}}}
\labfig{-1}
\end{figure}

The form of $\BL$ in \eq{x10} has the appealing feature that it has a positive definite imaginary part when $\Go$ has a positive imaginary part, corresponding to
fields that grow exponentially in time until one stops the experiment. Of course $\Gk$ and $\BGr$ also depend on frequency, but if we assume the
material is passive (not creating energy) then, by a direct extension
of the analysis in \cite{Welters:2014:SLL} (see also \cite{Milton:2007:MNS}), both $-\Gk(\Go)/\Go$ and $\Go\BGr(\Go)$ have positive imaginary parts when $\Go$ has a positive imaginary part.
As a consequence of this we can clearly express the solution as a minimization of fields in a variational principle \cite{Milton:2009:MVP, Milton:2010:MVP} by
making the Gibiansky-Cherkaev
transformation to equations similar to (4.4) and (9.2) in Part I. Indeed, recognizing that the acoustic (and elastodynamic and electrodynamic)
equations should be written
in this form so that one could apply the Gibiansky-Cherkaev transformation was the main contribution in \cite{Milton:2009:MVP}. In the typical case in a lossy material when $\Go$ is real, $\BGr(\Go)$ is real and strictly positive definite,
while $\Imag \Gk$ is positive, corresponding to the air or fluid having some compressive viscosity, we can still develop such variational principles
by first multiplying $\BL(\Bx)$ by $e^{i\vartheta}$ where $\Gvt$ is a small parameter chosen so that $e^{i\vartheta}\BL(\Bx)$ has a strictly positive definite imaginary part.
One normally thinks of liquids as having a shear viscosity, but not a compressive viscosity. However if the liquid is almost incompressible but has air bubbles
or other highly compressible (nonlossy) inclusions in it, then the liquid near the bubbles is necessarily sheared as they are compressed. So shear
viscosity is converted to compressive viscosity. In fact due to the incredible shearing near the bubbles, the effective compressive viscosity blows up as the
volume fraction of the bubbles decreases: see Section 11.4 of \cite{Milton:2002:TOC}. This shearing is then associated with enormous heating near the bubbles
which could be a contributing factor to sonoluminescence, where highly compressed bubbles emit light \cite{Brenner:2002:SBS}. Nonlinear effects, cavitation, and bubble
collapse are also important, with the temperatures reaching $10^4$ Kelvin. Nonlinear effects are even important
when analyzing the Minnaert resonance associated with bubble oscillations \cite{Ammari:2018:MRA}.
If both $\Go$ and $\Gk$ and $\BGr$ are real and
positive in one region, while $\BGr(\Go)$ is real and  strictly positive definite in that region, the variational principle requires one to use trial fields
that satisfy the field equations exactly in that region. The minimization variational principles also extend to scattering problems \cite{Milton:2017:BCP}.

Note that we are
always free to multiply $\BL$ by a constant that we may choose to be $\Go$ giving
\beq  \BL=\begin{pmatrix} -\Gk & 0 \\ 0 & \Go^2\BGr\end{pmatrix}, \eeq{x10a}
and assuming $\BGr$ is isotropic and constant, i.e., $\BGr(\Bx)=\Gr\BI$,
this is convenient for rewriting the equations in the form involving the resolvent in \eq{conve4} with $\underline{z}=\Gr\Go^2$.
We mention too that various other waves, such as the transverse electric and transverse magnetic equations of electromagnetism are governed by the
Helmholtz equation and thus the analysis here extends to them too.

\section{Elastodynamic equations at constant frequency}
\setcounter{equation}{0}
\labsect{3}

Beyond the  quasistatic regime, the elastodynamic equations at constant frequency $\Go$ take the form
\beq \begin{pmatrix}\BGs \\ \Div\BGs(\Bx)\end{pmatrix}=\BL\begin{pmatrix}-\Go\Grad \Bu \\ -\Go\Bu\end{pmatrix}-\bpm 0 \\ \Bf \epm,
\quad -i\Go\Bp(\Bx)=\Div\BGs(\Bx),
\eeq{15.8a}
where $\Bf$ is the complex body force density (assuming the actual body oscillate at the frequency $\Go$ so that they are given by
the real part of $e^{-i\Go t}\Bf$. If there are additional oscillatory sources of stress (such as due oscillatory thermal expansion or oscillatory swelling due to
humidity) then these can be included by taking their divergence and subtracting it from $\Bf$).  
Note that
we have put $\Grad\Bu$ in \eq{15.8a} rather than the strain, letting $\BCC(\Bx)$ do
the symmetrization (thus $\BCC(\Bx)$ acting on an antisymmetric matrix gives 0). We then have
\beq \BL(\Bx)=\begin{pmatrix}
-\BCC(\Bx)/\Go & 0 \\
0 & \Go\BGr(\Bx)\end{pmatrix},\quad \BGG_1(\Bk)=\BZ(\Bk),
\eeq{x7aa}
where $\BZ(\Bk)$, defined by \eq{x10}, just acts on the first index of the top element of the field. 

Now both $\BCC(\Bx)$ and the effective mass density term $\BGr(\Bx)$ may be frequency dependent and anisotropic. In passive media both
$-\BCC(\Go,\Bx)/\Go$ and  $\Go\BGr(\Go,\Bx)^{-1}$ have non-negative imaginary parts \cite{Milton:2007:MNS}, so that $\Imag\BL(\Bx)$
is positive semidefinite for all $\Bx$,  and this holds
even when $\Go$ is complex with a positive imaginary part (as one can see by a direct extension
of the analysis in \cite{Welters:2014:SLL}). Again if $\Go$ has a positive imaginary part or the material is lossy one can make
the Gibiansky-Cherkaev transformation \cite{Cherkaev:1994:VPC} and develop minimization variational principles \cite{Milton:2009:MVP, Milton:2010:MVP}.
In the typical case where $\Go$ is real and $\BGr(\Go,\Bx)$ is real
one can still develop the variational principles provided $-\Imag\BCC(\Go,\Bx)$ is strictly positive definite, at least in some regions.
Multiplying $\BL$ by $\Go$, and assuming $\BGr$ is isotropic and constant, i.e., $\BGr(\Bx)=\Gr\BI$ we obtain an $\BL$ such that one has equations involving the resolvent in \eq{conve4} with $\underline{z}=\Gr\Go^2$.

Unusual effects elastodynamic effects can occur in  metamaterials, as reviewed in \cite{Banerjee:2011:IMW, Srivastava:2015:EMD}. In
particular there could be off diagonal coupling terms in $\BL(\Bx)$, so that an acceleration causes a macroscopic stress, and a variation
in the strain rate influences the momentum. Local responses of this sort are captured in the local Willis equation
(see (2.4) in \cite{Milton:2006:CEM}), with
\beq \BL(\Bx)=\begin{pmatrix}
-\BCC(\Bx)/\Go & \BD(\Bx) \\
\BD(\Bx)^\dagger & \Go\BGr(\Bx)\end{pmatrix}.
\eeq{mbw}
The off diagonal terms are the Willis coupling terms \cite{Willis:1981:VRM, Willis:1981:VPDP, Willis:1997:DC} and, as will be
discussed in Section 7 of Part  III \cite{Milton:2020:UPLIII}, arise naturally in moving media. In Willis's original formulation $\BL$ (with its coupling terms)
was an operator, acting nonlocally in both space and time. His equations retain a symmetric stress field and a stress that only
depends on the symmetrized displacement gradient. Willis was the first one to propose this unexpected non-local coupling between
acceleration and stress, and momentum and strain rate. However, we will argue at the end of
Part IV \cite{Milton:2020:UPLIV}, that such nonlocal couplings can be
filtered out: they have no physically observable meaning (though theoretically and numerically they can be recovered
if one introduces ``eigenstrains'' that are uncorrelated with the microstructure \cite{Willis:2009:EER, Willis:2011:ECR}).
On the other hand, a non-local anisotropic density operator
does have a physical meaning in the context of ensemble averaged equations if we set
the coupling to zero, as we are free to do.

It was observed \cite{Milton:2006:CEM} that the local Willis equations were analogous to the bianisotropic equations of electromagnetism \cite{Serdyukov:2001:EBAM} and the form of the Willis equations was found to be preserved
under arbitrary spatial curvilinear coordinate transformations (see Appendix B and Appendix C of \cite{Milton:2006:CEM}).
Due to this, building upon the ideas of ``transformation optics'', it was suggested that materials with $\BL(\Bx)$ taking the form \eq{mbw}
might be useful for elasticity cloaking against small amplitude, time harmonic, incident elastic waves \cite{Milton:2006:CEM}. In fact,
under spatial transformations one can avoid inducing the coupling terms if one
allows for stress fields that are nonsymmetric and elasticity tensors that lose their
minor symmetries \cite{Brun:2009:ACI, Norris:2011:ECT, Vasquez:2013:SRP}. There are limitations to elastodynamic cloaking. Clearly if one wants to cloak
a large enough mass and moves the body far enough (one body length would suffice) then that mass will be felt through its inertial resistance.
An indepth study of limitations to elastodynamic cloaking is in \cite{Yavari:2019:NLE}.

           Of course the theory is no good without the support of examples
showing such local couplings. One  model was presented in \cite{Milton:2007:NMM} and simplified in \cite{Milton:2016:ETC}, reproduced here in \fig{1}
(see also Figure 1 in \cite{Koo:2016:AOM} and Figures 2 in \cite{Muhlestein:2017:EEW, Sieck:2017:OWC}).
The model in \cite{Milton:2007:NMM} has a $\BCC(\Bx)$ that breaks some of the usual symmetries,
thereby permitting the stress-field to be nonsymmetric). Further evidence in support of couplings
was provided by Willis \cite{Willis:2009:EER} who showed that an ensemble of laminate
geometries, randomly translated, had such a coupling in the long
wavelength limit, corresponding to local behavior. 
          Such a coupling has also been extended to acoustics \cite{Muhlestein:2016:MAH, Sieck:2017:OWC}, as mentioned in the previous section,
and to piezoelectricity with electro-momentum coupling \cite{Pernas:2020:SBC}.

Before the term ``transformation optics'' (and more generally ``transformation physics'' of various kinds)
was coined, Dolin \cite{Dolin:1961:PCT} had used it to
conceive objects that would be invisible to time harmonic applied fields,
being transformations of empty space;
Derrick, McPhedran, Maystre, and Nevier \cite{Derrick:1979:CGT} and Chandezon, Raoult, and Maystre \cite{Chandezon:1980:NTM}
used it to map doubly periodic
diffraction gratings to an equivalent material with a flat surface; Luc Tartar realized ``transformation conductivity'' could be applied to inverse problems
in conducting bodies, and remarked upon it to Kohn and Vogelius \cite{Kohn:1984:IUC}; it was used in section 8.5 of \cite{Milton:2002:TOC} to generate
equivalent classes of microstructures for which one could exactly solve for the fields and effective constants;
Lassas, Greenleaf, and Uhlmann \cite{Greenleaf:2003:ACC, Greenleaf:2003:NCI} used a singular
transformation, effectively stretching a point into a circular hole, to cloak conducting objects;
Leonhardt used ``transformation geometrical optics'' to cloak objects in the
geometric optics limit when the wavelength is extremely small \cite{Leonhardt:2006:OCM}.

\begin{figure}[!ht]
\centering
\includegraphics[width=0.7\textwidth]{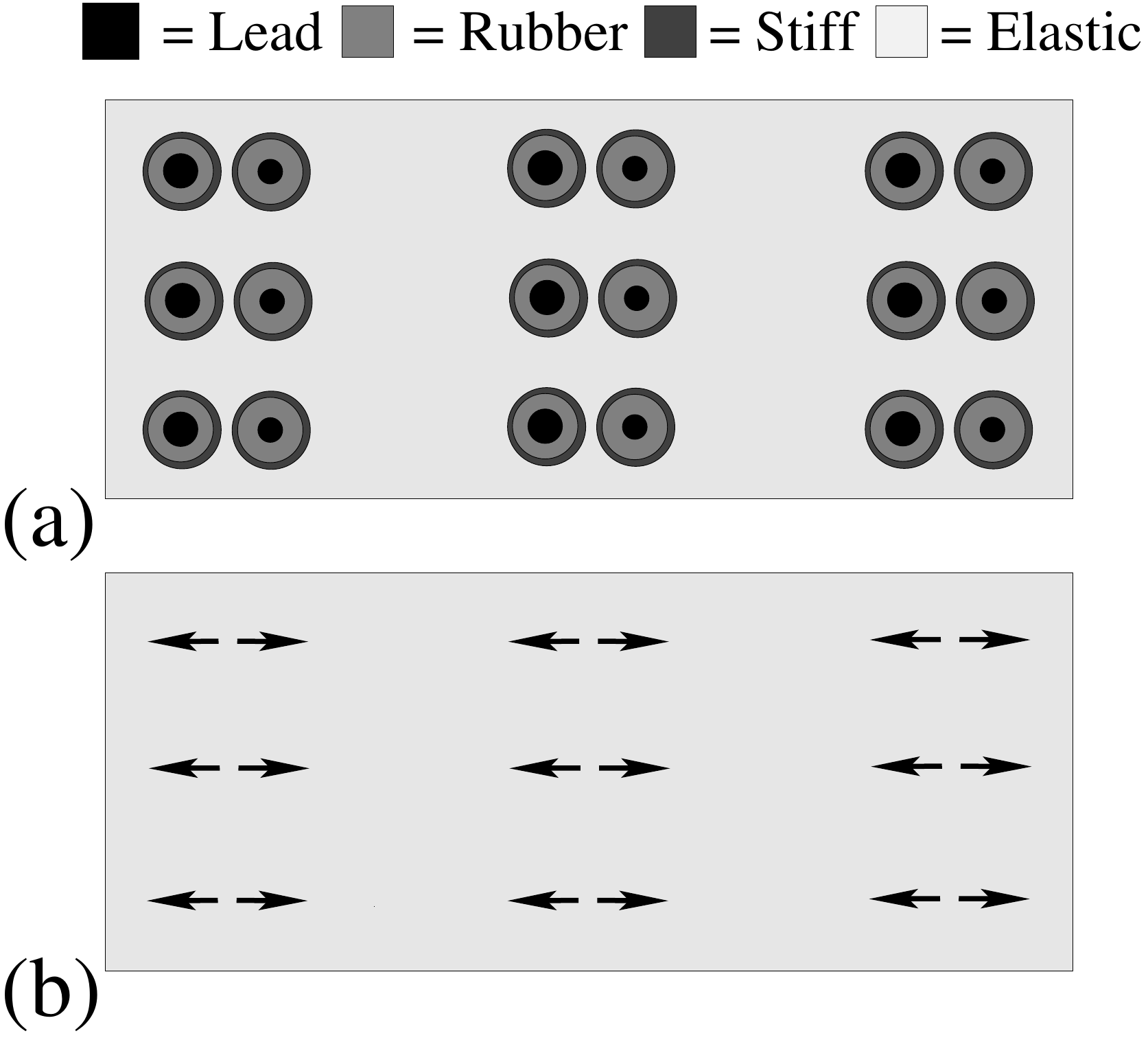}
\caption{A mechanism for producing a material with couplings where acceleration produces stress, and time varying strain produces momentum.
  The lead balls, surrounded by rubber, and coated by a shell
  of stiff material have a different amount of lead on the left and right sides of each pair, as shown in (a). As indicated by the work
  \protect{\cite{Auriault:1985:DCE, Auriault:1994:AHM}} and as experimentally and
  theoretically confirmed \protect{\cite{Sheng:2003:LRS, Liu:2005:AMP, Sheng:2007:DMD}}
  this can lead to the material in the shell on the left having a
  negative effective mass $-m$ and the material in the shell on the right having an almost equal and opposite positive effective
  mass $+m$, for a suitably tuned frequency. At this frequency when one accelerates the material back and forth
  these oscillating effective masses generate an array of oscillating force dipoles acting in the matrix as in (b) for one moment in time. Just like
  an array of electrical dipoles gives rise to an average polarization field, so too does an array of force dipoles give rise to an average stress field when the strain field is zero. Thus
  the time harmonic acceleration of the material gives rise to a time harmonic oscillating average uniaxial stress directed lengthwise when the average strain is
  zero, which is a characteristic feature when $\BL(\Bx)$ in \eq{15.8a} has couplings like in \eq{mbw}. Conversely, a uniaxial strain rate, pushes the mass pairs
  together and apart thus generating an oscillating momentum, because of their opposite effective masses.
  Even if the effective masses are not balanced, their net (monopolar) inertial force will be compensated by a net force on the inclusions from the surrounding medium.
  One still has an array of force dipoles that  create stress. Reproduced from Figure 1.2 in \protect{\cite{Milton:2016:ETC}}.
}
\labfig{1}
\end{figure}

An explicit model exhibiting (over a very narrow frequency band) a response governed by the local Willis equations, but with a nonsymmetric stress,
was constructed in \cite{Milton:2007:NMM}.
Nonsymmetric stress fields were introduced by the Cosserat brothers \cite{Cosserat:1968:FPC}. In particular macroscopic
nonsymmetric stresses can arise if there is interaction with some hidden spinning elements at the microscale, or more simply
if there are hidden mass elements with varying speeds of rotation \cite{Nassar:2018:DPL, Zhang:2020:AEM} or elements that simply
oscillate back and forth in a way that torque is transferred to the surrounding
structure \cite{Vasquez:2013:SRP}. In either case there is an interchanging of angular momentum between the hidden elements at
microscopic scale and the macroscopic scale. Then nonsymmetric stress fields are associated with the need to apply net torque
on the macroscopic material to stop it rotating. At fixed frequency one can build mass-spring models where, to account for this effect,
the springs are replaced by torque springs \cite{Vasquez:2013:SRP}. 

It is well known that a rotating disk hidden in a suitcase surprises
the person carrying the suitcase if they turn around a corner. Similarly, a metamaterial built from many such suitcases
stacked together will behave strangely. If the spinning disks inside the suitcases are weighted on one side (like often
happens in washing machines during the spin cycle if the clothes are not properly distributed) then there can be
a net vibration of the metamaterial, and conversely circular vibrations of the metamaterial can induce the
spinning disks to rotate without the need to power their motion in some nonmechanical way. One might call
such metamaterials ``spinetic'' due to the obvious similarity with magnetic materials (caused by the magnetic field associated
with charged particles with intrinsic spin, or rotating charged particles). Thus, the action of making the disks
spin (by vibration or other means) would be spinetizing it to reach a spinitized state.


Motivated by the work of the  Cosserat brothers, Cartan suggested a modification to
Einsteins general relativity, incorporating torsion \cite{Cartan:1955:MAC}. Later the intrinsic spin of particles was discovered and
this provided additional grounds for the necessity of having nonsymmetric stress fields.
Among other candidates, torsion might explain the dark matter and dark energy in the universe: see \cite{Milton:2020:PED} and references therein, that include
references to reviews of the many torsion theories. 

We stress, and this applies to the time dependent elastodynamic equations as well, that anisotropic, frequency dependent density,
and the associated terms coupling the stress to the acceleration and the momentum to the strain rate at constant frequency are only valid concepts when the internal and macroscopic vibrations are small compared to say
the size of the unit cell of periodicity in a metamaterial. For larger vibrations it is probably a better approximation to use the classical elastodynamic equations
as then the internal masses will get dragged along with the rest of the unit cell.
\section{Brinkman-Stokes-Darcy and compressible Oseen flow equations
  with perturbations at constant frequency}
\setcounter{equation}{0}
\labsect{4}
The equations governing flow of an incompressible in a porous medium are often taken to be the unsteady Brinkman equations \cite{Tumuluri:2018:CGS},
which at constant frequency take the form:
\beq \bpm i[\Grad\Bv+(\Grad\Bv)^T]/2 \\ i\Bv \epm=\BL\bpm \BGs \\ \Div\BGs \epm +\BL\bpm 0 \\ \Bf \epm, \eeq{bsd1}
where $\Bv(\Bx)$ is the macroscopic fluid velocity field and $\BGs(\Bx)=\BGs_s(\Bx)-P(\Bx)\BI$ is the stress, in which $\BGs_s$ is the shear stress,
$P$ is the pressure, and the source $\Bf$ is a time harmonic body force, such as an electrical force if the fluid is charged
or has electric dipoles and is in the presence of  time harmonic electric and/or magnetic fields or their gradients.
The differential constraints on the fields are accounted for by taking 
\beq \BGG_1(\Bk)=\BI-\BGG_2(\Bk), \quad \BGG_2(\Bk)=\bpm \BD(i\Bk) \\ \BI \epm \bpm \BD^\dagger(i\Bk)\BD(i\Bk)+\BI\epm^{-1}\bpm \BD^\dagger(i\Bk) & \BI \epm,
  \eeq{bsdgam}
  where the action of $\BD(i\Bk)$ and its adjoint $\BD^\dagger(i\Bk)$ on a vector $\Bm$ and symmetric matrix $\BM$ are given by
  \beq \BD(i\Bk)\Bm=\frac{i}{2}(\Bm\otimes\Bk+\Bk\otimes\Bm),\quad \BD^\dagger(\Bk)\BM=-i\Bk\cdot\BM.
  \eeq{bdsgam0}
Then, allowing the porous medium to be anisotropic, we have
\beq \BL(\Bx)=\bpm i\BCV(\Bx) & 0 \\ 0 & -[\Go\BGr(\Bx)+i\Gn(\Bk(\Bx))^{-1}]^{-1} \epm,
\eeq{bsd2}
where $\BCV(\Bx)$ is a real fourth order viscosity tensor, positive definite for all $\Bx$ on the space of symmetric matrices,
that relates the shear strain rate to the shear stress $\BGs_s$;
$\Bk(\Bx)$ is the matrix valued fluid permeability tensor; $\Gn$ is the dynamic shear viscosity of the
fluid; and $\BGr(\Bx)$ is the effective fluid density that can be anisotropic. Note that $\BCV(\Bx)$ annihilates symmetric tensors proportional to $\BI$ and
only produces trace free symmetric tensor fields, i.e.,
\beq  \BGL_h\BCV=\BCV\BGL_h=0, \quad \BGL_s\BCV=\BCV=\BCV\BGL_s, \eeq{bsd3}
in which $\BGL_h$, and $\BGL_s$ are the projections onto matrices proportional to the identity matrix and 
tracefree symmetric matrices, respectively. As a consequence, \eq{bsd1} implies that $\Div\Bv=0$
(the fluid is incompressible) and $\BCV\BGs=\BCV\BGs_s$. So we can rewrite \eq{bsd1} as
\beq \Div\BGs=\Div\BCV^{-1}[\Grad\Bv+(\Grad\Bv)^T]/2-\Grad P=[-i\Go\BGr(\Bx)+\Gn[\Bk(\Bx)]^{-1}]\Bv-\Bf, \quad \Div\Bv=0, \eeq{bsd4}
implying the equation
\beq -i\Go\BGr(\Bx)\Bv=\Div\BCV^{-1}[\Grad\Bv+(\Grad\Bv)^T]/2-\Grad P-\Gn[\Bk(\Bx)]^{-1}\Bv +\Bf, \eeq{bsd5}
that easily can be related to the standard Brinkman equation. For instance, the term on the left hand side
of \eq{bsd5} is associated with the inertial term $\BGr\Md \Bv/\Md t$. 
The time harmonic Brinkman equations can be seen as a blend of the time harmonic Stokes equations for creeping flow (obtained in the limit
where the drag term $\Gn[\Bk(\Bx)]^{-1}\Bv$ goes to zero) and the time harmonic
Darcy equations (obtained in the limit where the shear stress contribution $\Div\BCV^{-1}[\Grad\Bv+(\Grad\Bv)^T]/2$ goes to zero).
While the Brinkman equations were first derived using heuristic arguments, and thus with uncertain foundations,
there is a body of theoretical and experimental results for
supporting them when the solid volume fraction is low: see \cite{Durlofsky:1987:ABE} and references therein. One advantage of
Brinkman's equations is that being second order in the velocity $\Bv$, as opposed to Darcy's equation which is first order,
one can have nonslip conditions for flow in a porous medium around a solid, nonporous, object. We see that the tensor
$\BL$ in \eq{bsd2} has a positive definite imaginary part, allowing for the development of Cherkaev-Gibiansky type variational principles.

In the case of compressible fluids one can
keep the form \eq{bsd1} and \eq{bsd2}, replacing $i\BCV(\Bx)$ by $\BCS(\Bx)$ where $\BCS(\Bx)$ is some fourth order compliance tensor that has an
imaginary part that is positive definite on the space of symmetric matrices (thus, in particular, $\BCS(\Bx)\BI \ne 0$). The imaginary part of $\BCS(\Bx)$
is then related to the viscous dissipation. The equations become
\beq \bpm \BGs \\ \Div\BGs \epm=\BL^{-1}\bpm i\Grad\Bv \\ i\Bv \epm-\bpm 0 \\ \Bf \epm,\quad \BGG_1(\Bk)=\BZ(\Bk),
\eeq{bsd5com}
where the much simpler form of $\BGG_1(\Bk)$ results from letting $\BL^{-1}$ do the symmetrization of $\Grad\Bv$.

Constant frequency perturbations to compressible Oseen viscous (low Reynolds number) flow around an object moving with constant velocity $\BU$ in
a frame of reference moving with the object also take the form \eq{bsd5com} where $\Bv$, the velocity perturbation, satisfies $\Bv=0$ at the surface of
the object, $\Bf$ is the oscillatory forcing, and  
\beq \BL^{-1}=\bpm \BCC & 0 \\ \BU\cdot & -\Go\BGr(\Bx) &\epm, \eeq{bsd6}
in which the isotropic fourth order complex elasticity tensor takes the form
\beq \BCC=(\Gk-i\Go\Gn_B)\BGL_h/3-2i\Go\Gn\BGL_s, \eeq{bsd7}
in which $\Gk$ is the bulk modulus and $\Gn_B$ and $\Gn$ are the compressive viscosity and shear viscosity. These viscosities depend on the temperature,
which has a contribution from the heat generated by the steady flow of fluid around the object: Thus one can expect that $\Gn_B$ and $\Gn$ will
depend on $\Bx$.

\section{Navier-Stokes incompressible fluid equations with constant frequency perturbations}
\setcounter{equation}{0}
\labsect{5}
Suppose that $\Bv(\Bx,t)$ is the fluid velocity that satisfies the Navier-Stokes equations
in an incompressible fluid having constant density $\Gr$:
\beq \Gr\frac{\Md \Bv}{\Md t}+\Gr(\Bv\cdot\Grad)\Bv-\Div\Gn\Grad\Bv=-\Grad P +\Bf,\quad \Div\Bv=0,
\eeq{ns1}
where $\Gn$ is the  shear viscosity.
Now replace each field $\Bv(\Bx,t)$, $P(\Bx,t)$, and $\Bf(\Bx,t)$, by $\Bv(\Bx)+\Ge\Real[e^{-i\Go t}\Bv'(\Bx)]$, $P(\Bx)+\Ge\Real[e^{-i\Go t}P'(\Bx)]$,
and $\Bf(\Bx)+\Ge\Real[e^{-i\Go t}\Bf'(\Bx)]$, where $\Bv(\Bx)$ and $P(\Bx)$ are steady state solutions to the Navier-Stokes equations with
body force  $\Bf(\Bx)$. The force $\Bf$ may be due to gravity, while the oscillatory part $\Bf'$ may be due to oscillatory
accelerations of the system (acting like an oscillatory gravitational force due to Einstein's equivalence principle)
or due to oscillatory electric fields gradients if the fluid contains electrical dipoles. 
By substituting these in \eq{ns1} to first order in $\Ge$ we get
\beq -i\Go \Gr\Bv'+ \Gr(\Bv'\cdot\Grad)\Bv+ \Gr(\Bv\cdot\Grad)\Bv'-\Div\Gn\Grad\Bv'=-\Grad P' +\Bf',\quad \Div\Bv'=0.
\eeq{ns2}
Due to heating of the steady state fluid component, caused by the viscous terms, the shear viscosity $\Gn$ as it depends on the
local temperature will also be a function of $\Bx$, $\Gn=\Gn(\Bx)$.  We now rewrite the equations for the time harmonic perturbed fields as:
\beq \bpm \BGs_s' -P'\BI\\ -P' \\ \Div(\BGs_s'-P'\BI)\epm
=\BL\bpm \Grad\Bv' \\ \Div\Bv' \\ \Bv' \epm+\bpm 0\\ 0\\ \Bf' \epm,
\eeq{ns3}
with
\beq \BL=\bpm 2\Gn\BGL_s +\infty\BGL_h & 0 \\
\Gr\Bv\cdot &  -i\Go\widetilde{\BGr} \epm,\quad \widetilde{\BGr}=\Gr(\BI+i(\Grad\Bv)^T/\Go),
\quad \BGG_1(\Bk)=\BZ(\Bk),
\eeq{ns4}
in which $\BGL_s$ and $\BGL_h$ are the projections onto trace free symmetric matrices, including $\BGs_s'$, and matrices proportional
to $\BI$, respectively. Here as in the Oseen equations treated in Part I,
$\infty$ should be considered to be a large parameter that we 
let approach infinity. In this limit $\Div\Bv'$ is forced to zero, corresponding
to the incompressibility of the flow, while $P'(\Bx)$ is unconstrained, except through the $\Grad P'$ term
in \eq{ns3}. Of course stationary perturbations to the Navier-Stokes equations are obtained by letting $\Go\to 0$,
thus replacing $-i\Go\widetilde{\BGr}$ with $\Gr(\Grad\Bv)^T$ in $\BL(\Bx)$.

\section{Time harmonic linear thermoacoustic equations}
\setcounter{equation}{0}
\labsect{6}

I found it difficult to find appropriate formulations of the time harmonic linear thermoacoustic equations in
the published literature. The main source here is \cite{COMSOL:2013:AMU}, equations (7-5), page 286, see also \cite{Pierce:1981:AIP},
built upon in Section 1.8 of \cite{Milton:2016:ETC}, giving
\beq
\begin{pmatrix}
i\BGs \\
i\Div\BGs \\
i\Bq \\
i\Div\Bq
\end{pmatrix}
=\BL\begin{pmatrix}
\Grad\Bv \\
\Bv \\
\Grad \Gt/T_0 \\ \Gt/T_0\end{pmatrix}-\bpm 0 \\ i\Bf \\ 0 \\ i\BQ \epm,
\eeq{15.8aa}
where the real parts of  $e^{-i\Go t}\BGs$, $e^{-i\Go t}\Bq$, $e^{-i\Go t}\Gt$, $e^{-i\Go t}\Bv$, $e^{-i\Go t}\Bv=\Bf$, and $e^{-i\Go t}\BQ$ are the time harmonic
stress, heat current, temperature fluctuations, fluid velocity,  volume force density, and heat source density,
while $T_0$ is the constant background temperature.
One then has that
\beqa
\BL(\Bx)& = & \begin{pmatrix}
i\BCD(\Bx)+\frac{\BI\otimes\BI}{\Go\Gb_T} & 0 & 0 & \frac{-i\Ga_0T_0\BI}{\Gb_T}  \\
0 & -\Go\Gr_0 & 0 & 0  \\
0 & 0 & i\BK(\Bx)T_0 & 0 \\
\frac{i\Ga_0T_0\BI\cdot}{\Gb_T} & 0 & 0 & \Go\frac{\Ga_0^2T_0^2}{\Gb_T}-\Go\Gr_0C_pT_0 
\end{pmatrix},\nonum
\BGG_1(\Bk) & = & \bpm \BZ(\Bk) & 0  \\ 0 & \BZ(\Bk) \end{pmatrix},
\eeqa{x77}
where $\BZ(\Bk)$ is given by \eq{x10} and
$\BCD=\Gn_B\BGL_h/3+2\Gn\BGL_s$ is the isotropic fourth order viscosity tensor, $\Gn_B(\Bx)$ and $\Gn(\Bx)$ being
the compressive and shear viscosities of the fluid, $\Gb_T(\Bx)$ is the isothermal compressibility, $C_p(\Bx)$ is the heat capacity at constant pressure;
$\Ga_0$ is the coefficient of thermal expansion at constant pressure; and $\BK(\Bx)$ is the thermal conductivity tensor. Both the background density $\Gr_0$
and the background temperature $T_0$ are assumed to be constant. Note that the first $\BZ(\Bk)$ in $\BGG_1(\Bk)$ acts on the first index of
$(matrix, vector)$ fields while the second $\BZ(\Bk)$ acts on $(vector, scalar)$ fields. Interestingly, thermoacoustic engines hold the
promise of converting waste heat into
energy \cite{Flitcroft:2013:UTE}.

\section{Surface waves and associated internal waves}
\setcounter{equation}{0}
\labsect{7}
Many planar surface waves are governed by equations of the form \eq{ad1}. Consider, for example,
the Love surface waves in elasticity. When propagating along the $x_1$-axis at constant frequency
$\Go$ along the surface $x_3=0$, the only nonzero component of the displacement field is the vertical
component $u_3(x_1,x_3,t)=u(x_3)e^{ik_1x_1-\Go t}$, where $u(x_3)$ satisfies
\beq \bpm \Gj \\ \Md \Gj/\Md x_3 \epm =\BL\bpm \Md u /\Md x_3 \\ u \epm, \quad \BL(x_3)=\bpm \Gm(x_3) & 0 \\ 0 & k_1^2\Gm(x_3)-\Go^2\Gr(x_3) \epm,
\quad \BGG_1(k_3)=\frac{1}{k_3^2 +1}\bpm k_3^2 & ik_3 \\ -ik_3 & 1 \epm,
\eeq{sw.1}
in which $\Gj$ is a shear component, and the shear modulus $\Gm$ and density $\Gr$ only vary in the vertical $x_3$ direction.

More generally we can consider solutions to \eq{ad1} with $\BL$ only depending on $x_3$ and $\BJ(\Bx)$, $\BE(\Bx)$
and $\Bs(\Bx)$ having the form
\beq \BJ(\Bx)=\BJ^0(x_3)e^{ik_1x_1-\Go t},\quad \BE(\Bx)=\BE^0(x_3)e^{ik_1x_1-\Go t},\quad \Bs(\Bx)=\Bs^0(x_3)e^{ik_1x_1-\Go t}.
\eeq{sw.2}
Then the equations reduce to
\beq \BJ^0(x_3)=\BL(\Bx_3)\BE^0(x_3)-\Bs^0(x_3), \quad \BGG^0\BE^0=\BE^0,\quad \BGG^0\BJ^0=0,\quad \BGG^0(k_3)=\BGG_1((k_1,0,k_3)),
\eeq{sw.3}
where $k_1$ is to be treated as a constant in this last expression. 
When one has a sharp discontinuity in material properties at a surface, these equations need to be interpreted in their weak
form. However we remark that there is no need to have a sharp surface. One looks for solutions where the fields vanish as $|x_3|$ goes
to infinity. A good example of internal waves are the enormous (albeit nonlinear) internal ocean waves \cite{Mercier:2013:LSR}.
Of course these one dimensional equations are trivial to solve numerically and they are only introduced to make a connection with the equations \eq{ad1}.

\section{ Electromagnetic equations at constant frequency with free current sources}
\setcounter{equation}{0}
\labsect{8}

At constant frequency Maxwell's equations take the form:
\beq \bpm -\Grad \times \Bh \\ \Bh \epm=\BL \bpm i\Be \\ i\Grad \times \Be \epm -\bpm \Bj_f \\ 0\epm,
\eeq{15.13}
where $\Bj_f$ is the free current source assumed to oscillate with frequency $\Go$ (not to be confused with the induced current source
associated with time variations of the electric displacement field). We then have
\beq \BL(\Bx)=\begin{pmatrix} \Go\BGve(\Bx) & 0 \\ 0 & -[\Go\BGm(\Bx)]^{-1} \end{pmatrix}, \quad
\BGG_1(\Bk)=\bpm \BI \\ i\BGn(\Bk) \epm [\BI-\BGn(\Bk)\BGn(\Bk)]^{-1}\bpm \BI & i\BGn(\Bk) \epm.
\eeq{x13}
Here the action of $\BGn(\Bk)$ on vector $\Ba$ gives $\BGn(\Bk)\Ba=\Bk\times\Ba$. Thus $\BGn$ has indices $\Gn_{ijk}$ that are $1$ or $-1$
according to whether $ijk$ is an even or odd permutation of $1,2,3$, and so $\BGn(\Bk)$ is an antisymmetric matrix. The magnetic permeability tensor $\BGm(\Bx)$ and
electric permeability tensor $\BGve(\Bx)$ are typically anisotropic and complex. The matrix inverse in \eq{x13} is easily computed as
upon noting that $\BGn(\Bk)\BGn(\Bk)=\Bk\otimes\Bk-k^2\BI$ it becomes the inverse of a linear combination of the identity matrix and a rank one matrix. 
That inverse is given by the Sherman-Morrison formula and we obtain
\beq \BGG_1(\Bk)=\frac{1}{k^2+1}\bpm \BI \\ i\BGn(\Bk) \epm [\BI+\Bk\otimes\Bk]\bpm \BI & i\BGn(\Bk) \epm. \eeq{x13simp}

The perturbed constant frequency elastodynamic equations at constant
frequency $\Go$ are of particular importance for magnetotellurics
where one probes the subsurface conductivity of the earth or ocean
\cite{Thiel:2008:MIM}. These take the form \eq{15.13} with
$\Bj_f=-(\BGs-\BGs^p)\BE^p$ where $E^p$ and $\BGs^p$ are the given
primary (unperturbed) electric field and complex conductivity (admittance),
while $\BGs-\BGs^p$ is the perturbation to the complex conductivity
and $\Ge$ is a small scaling parameter. Thus the total field is
$\BE_p+\BE$.
The magnetotelluric source field $\BE^p$ is usually taken to be the
electric field associated with a vertically propagating planar electromagnetic
wave, and $\BGs^p$ is taken to be $-i\Go\Gve_0$ above the earth's surface
and constant below it with a nonzero real part (ensuring square integrability
of the field $\BE(\Bx)$). 
Measurements are made at the earth's surface of the complex
valued magnetotelluric impedance tensor $\BZ(\Go,\Bx_0)$ and
tipper $\BK(\Go,\Bx_0)$ defined by
\beq  \bpm e_1(\Bx_0) \\ e_2(\Bx_0) \epm=\underbrace{\bpm Z_{11}(\Go,\Bx_0) & Z_{12}(\Go,\Bx_0) \\ Z_{21}(\Go,\Bx_0)
  & Z_{22}(\Go,\Bx_0)\epm}_{\BZ(\Go,\Bx_0)}\bpm h_1(\Bx_0) \\ h_2(\Bx_0) \epm, \quad
h_3(\Bx_0)=\underbrace{\bpm K_{1}(\Go,\Bx_0) & K_2(\Go,\Bx_0) \epm}_{\BK(\Go,\Bx_0)}\bpm h_1(\Bx_0) \\ h_2(\Bx_0)\epm
\eeq{mt0}
where $\Bx_0=(x_1,x_2,0)$ are coordinates at the earth's surface. Given
these measurements, the objective is to solve the inverse problem
of recovering $\BGs(\Bx)$: see \cite{Kordy:2016:MII} and references therein.

In some metamaterials there could be off diagonal
coupling terms in $\BL(\Bx)$ in \eq{x13}.  These are the bianisotropic coupling terms \cite{Serdyukov:2001:EBAM}, which are the
direct analogs for electromagnetism of the coupling terms in \eq{mbw}. They also arise naturally in moving media as we will see in Section 7 of
Part III \cite{Milton:2020:UPLIII}.
Analogously, to the effective density in \eq{x10}, the effective magnetic permeability can be
frequency dependent or negative even if the material is nonmagnetic having the same magnetic permeability as free space.
Schelkunoff and Friis \cite{Schelkunoff:1952:ATP} and Lagarkov et.al. \cite{Lagarkov:1997:RPB} realized one could get artificial
magnetism with arrays of split rings (see also \cite{Vinogradov:1997:RPB}). Experiments
showing negative magnetic permeability were performed by Lagarkov et.al. \cite{Lagarkov:1997:RPB}. Materials with both negative permittivity and permeability,
and hence a negative refractive index, as studied by Veselago \cite{Veselago:1967:ESS},
were experimentally realized by Shelby, Smith, and Schultz \cite{Shelby:2001:EVN}.  These materials have
the curious property that the wave crests move opposite to the group velocity, i.e., opposite to the direction
of energy flow. Waves with these properties were studied as early as 1904 \cite{Lamb:1904:GV, Schuster:1904:ITO}. 
Bouchitt{\'e}, Felbacq, and Schweizer \cite{Bouchitte:2004:HNR, Bouchitte:2010:HME} made rigorous analyses using homogenization theory
to support the claims. The link between artificial magnetic permeability in split ring geometries and anisotropic effective density
for antiplane elasticity in split ring cylinders was recognized by Movchan and Guenneau \cite{Movchan:2004:SRR}.

The form of $\BL$ in \eq{x10} has once again the appealing feature that it has a positive definite imaginary part when $\Go$ is real or has a positive imaginary part,
as for passive materials both $\Go\BGm(\Bx)$ and $\Go\BGve(\Bx)$ have non-negative imaginary parts \cite{Welters:2014:SLL}. If they are both positive definite one
has minimization variational principles, and these variational principles can still be developed, by multiplying $\BL$ by $e^{i\vartheta}$ 
if there are no magnetic losses, i.e. if $\BGm(\Bx)$ is real but positive definite \cite{Milton:2009:MVP, Milton:2010:MVP}.
Even if $\BGm(\Bx)$ and $\BGve(\Bx)$ are both real, but positive definite, then we can still develop the 
variational principles for frequencies $\Go$ having a positive imaginary part, i.e. having fields that grow exponentially
until measurements are taken. Similar manipulations can of course be done for the time harmonic elastodynamic and acoustic equations \eq{x7aa} and \eq{x10}.
in lossy media. Multiplying $\BL$ by $\Go$, and assuming $\BGve(\Bx)$ is constant and isotropic,
i.e. $\BGve(\Bx)=\Gve\BI$ we obtain an $\BL$ such that one has equations involving the resolvent in \eq{conve4} with $\underline{z}=\Gve\Go^2$.
Equivalently, thanks to the symmetry of Maxwell's equations (without sources) when we interchange $\Be$ and $\BGve$ with  $\Bh$ and $\BGm$,
we can alternatively make the more realistic assumption that $\BGm(\Bx)$ is constant and isotropic, i.e. $\BGm(\Bx)=\Gm\BI$, and then
we can obtain a resolvent with  $\underline{z}=\Gm\Go^2$.

An interesting side remark is that one can develop network models for
time harmonic electromagnetism that are analogous to mass-spring models
for time harmonic elastodynamics \cite{Milton:2009:EC, Milton:2010:HEC}. 
A mass-spring model can be viewed as
concentrated masses, say taking the form of rigid (infinite stiffness)
spheres, connected by
springs, say taking the form of cylindrical tubes of elastic material
having negligible mass, and surrounded everywhere by void --- namely
a material with zero stiffness and zero mass. For electromagnetism
there are two types of analogous network models: M-circuits and
E-circuits. An M-circuit (or E-circuit)
is a network of triangular prisms filled with material
having $\Gm\ne 0$ and $\Gve=0$ (or, respectively, $\Gve\ne 0$ and $\Gm=0$)
connected along their edges by cylinders having 
$\Gve\ne 0$ and $\Gm=0$ (or, respectively,  $\Gm\ne 0$ and $\Gve=0$)
surrounded by material having $\Gm=\infty$ and $\Gve=0$
(or, respectively,  $\Gm=0$ and $\Gve=\infty$). Materials
with $\Gve=0$ (an ENZ material) or $\Gm=0$ (an MNZ material),
or with $\Gve=\infty$ or $\Gm=\infty$, are not prohibited at a single frequency
and have some unexpected electromagnetic
properties \cite{Silveirinha:2006:TEE, Marcos:2015:MNZ}.

\section{Quasiperiodic solutions in lossy periodic media with quasiperiodic sources}
\setcounter{equation}{0}
\labsect{20}
Consider the time harmonic wave equations in an $\GO$-periodic lossy medium (where $\GO$ is the unit cell of periodicity)
with a quasiperiodic source term taking the form
\beq \Bs(\Bx)=e^{i\Bk_0\cdot\Bx}\Bs_0(1+\Ga_f(\Bx)), \eeq{qps.1}
where the wavevector $\Bk_0$ is any real vector, $\Bs_0$ is any complex constant vector, and the fluctuating scalar component $\Ga_f(\Bx)$ is $\GO$-periodic
with zero average value over $\GO$. The physical source is the real part of $e^{-i\Go t}\Bs(\Bx)$.
Any general  quasiperiodic source term can be regarded as a superposition of
source terms of the form \eq{qps.1}, as can be seen by taking $1+\Ga_f(\Bx)$ to be an $\GO$-periodic array of delta function, modulated
by $e^{i\Bk_0\cdot\Bx}$ times the volume of $\GO$. The solutions to \eq{ad1} take the related form:
\beq \BE(\Bx)=e^{i\Bk_0\cdot\Bx}(\BE_0+\BE_f(\Bx)),\quad \BJ(\Bx)=e^{i\Bk_0\cdot\Bx}(\BJ_0+\BJ_f(\Bx)),
\eeq{qps.2}
where the fluctuating components $\BE_f(\Bx)$ and $\BJ_f(\Bx)$ have zero average value over the unit cell of periodicity. Since $\BE_0$ and $\BJ_0$
are linearly related to $\Bs_0$ it seems natural to define ``effective tensors'' $\BL_*^E$ and $\BL_*^J$ via
\beq \BE_0=\BL_*^E\Bs_0,\quad \BJ_0=\BL_*^J\Bs_0. \eeq{qps.3}
These ``effective tensors'' depend nonlinearly on $\Bk_0$, $\BL(\Bx)$, but linearly on $\Ga_f(\Bx)$. Their utility remains to be assessed. They can be
obtained from the associated periodic Green's functions in the
inhomogeneous medium $\BL(\Bx)$. This interpretation does not
help when one is considering just one fixed modulation function
$\Ga_f(\Bx)$ (perhaps constant in each phase in a multiphase medium)
as these  Green's functions are difficult to compute and 
not translationally invariant. As the loss
goes to zero and $\Bk_0$ approaches a value on the dispersion relation $\Bk(\Go)$ then the norms of $\BL_*^E$ and $\BL_*^J$ both diverge to $\infty$.  

\section{Source free time harmonic multielectron Schr{\"o}dinger equation}
\setcounter{equation}{0}
\labsect{9}

We assume a time dependence $e^{-iEt/\hbar}$ where $E$ is the energy
and $\hbar$ is Planck's constant divided by $2\pi$. The time harmonic multielectron Schr{\"o}dinger equation is 
equivalent to the time harmonic acoustic equation in a multidimensional space, with symmetry requirements on the solution.
It can be written in the same form, with the
(generally complex valued) wavefunction $\psi(\Bx)$ playing the role of the pressure:
\beq \begin{pmatrix}\Bq \\\Div\Bq\end{pmatrix}=\BL\begin{pmatrix}\Grad \psi \\ \psi\end{pmatrix},
\eeq{15.16} 
where $\psi(\Bx)$ is the complex valued wavefunction. Here $\Bx$ lies in a multidimensional space $\Bx=(\Bx_1,\Bx_2,\ldots,\Bx_N)$ where following,
for example, \cite{Parr:1994:DFT}, each $\Bx_i$ represents a pair $(\Br_i,s_i)$ where $\Br_i$ is a three dimensional vector
associated with the position of electron $i$ and
$s_i$ denotes its spin (taking discrete values $+1/2$ for spin up or $-1/2$ for spin down). Accordingly, $\Grad$ represents 
\beq \Grad=\bpm \Grad_1 \\ \Grad_2 \\ \vdots \\ \Grad_N\epm,\quad\text{where}~~\Grad_\Ga=\bpm \Md/\Md \{\Br_\Ga\}_ 1 \\  \Md/\Md\{\Br_\Ga\}_2 \\ \Md/\Md\{\Br_\Ga\}_3 \epm.
\eeq{15.17}
For the time harmonic multielectron Schr{\"o}dinger equation we have
\beq \BL=\begin{pmatrix}-\BA & 0 \\ 0 & E-V(\Bx)\end{pmatrix},\quad
\BGG_1(\Bk)=\frac{1}{k^2+1}\begin{pmatrix}
\Bk\otimes\Bk & i\Bk \\
-i\Bk^T & 1
\end{pmatrix}=\frac{{\BD(i\Bk)}{\BD(i\Bk)}^\dagger}{k^2+1}\quad\text{with  }\BD(\Grad)=\bpm \Grad \\ 1 \epm,
\eeq{x17}
where $V(\Bx)$ is the potential and $\BA$ in the simplest approximation is $\hbar^2\BI/(2m)$ in which $m$ is the
mass of the electron, but it may take other forms to take into account the reduced mass of the
electron, or mass polarization terms, due to the motion of the atomic nuclei. Here $\Bk$ represents the Fourier coordinate
\beq
\Bk = \bpm \Bk_1 \\ \Bk_2 \\ \vdots \\ \Bk^N\epm,\quad\text{where}
~~\Bk_\Ga=\bpm \{\Bk_\Ga\}_1 \\  \{\Bk_\Ga\}_2 \\ \{\Bk_\Ga\}_3 \epm.
\eeq{x18}
As electrons are fermions, in solving \eq{x17} one requires that $\psi(\Bx)$ have the symmetry properties:
\beq \psi(\Bx_1,\ldots,\Bx_j,\ldots,\Bx_k,\ldots,\Bx_N) =
-\psi(\Bx_1,\ldots,\Bx_k,\ldots,\Bx_j,\ldots,\Bx_N),
\eeq{sy1}
with $j\ne k$, implying that $q(\Bx)$ has  the symmetry properties:
\beqa
\Bq_j(\Bx_1,\ldots,\Bx_j,\ldots,\Bx_k,\ldots,\Bx_N) &=&
-\Bq_k(\Bx_1,\ldots,\Bx_k,\ldots,\Bx_j,\ldots,\Bx_N),\nonum
\Bq_m(\Bx_1,\ldots,\Bx_j,\ldots,\Bx_k,\ldots,\Bx_N) &=&
-\Bq_m(\Bx_1,\ldots,\Bx_k,\ldots,\Bx_j,\ldots,\Bx_N),
\eeqa{sy2}
if $m\neq j$ and $m\neq k$.

\section{Time harmonic multielectron Schr{\"o}dinger equation with sources}
\setcounter{equation}{0}
\labsect{10}

Sources in the time harmonic multielectron Schr{\"o}dinger equation naturally arise in the context of perturbations, i.e. when
we slightly perturb $V(\Bx)$. The Schr{\"o}dinger equation \eq{15.16} with $\BL$ and $\BGG_1$ given by \eq{x17} has of course the
trivial solution $\psi(\Bx)=0$,
and to get nontrivial solutions $E$ must be such that $\BA\BGG_1\BA$ does not have an inverse, i.e. $E$ is in the spectrum of the
resolvent. We assume that $\BE$ corresponds to a single (nondegenerate) bound state with a square integrable  $\psi(\Bx)$,
so that $\BA\BGG_1\BA$ has a one dimensional null space. The wavefunction  $\psi(\Bx)$, satisfying \eq{15.16}, can be normalized so that
\beq \int \psi(\Bx)\overline{\psi(\Bx)}\, d\Bx=1, \eeq{p1}
where by the integral over $\Bx$ we imply not only an integral over $\Br=(\Bx_1,\Bx_2,\ldots,\Bx_N)$ over $\mathbb{R}^N$, but also a sum over the spin states
$s=(s_1,s_2,\ldots,s_N)$. The constraint \eq{p1} naturally follows from the interpretation of $\psi(\Bx)\overline{\psi(\Bx)}$ as a probability density.

Let $\Ge V'(\Bx)$ be the perturbation to $V(\Bx)$. We rewrite \eq{15.16} and \eq{x17} in the more conventional form:
\beq E\psi=H\psi,\quad\text{with}\quad H=-\Div\BA\Grad +V, \eeq{p1a}
involving the Hamiltonian operator $H$. Following the standard perturbation analysis and
replacing $E$, $V$, and $\psi$ by $E+\Ge E'$, $V+\Ge V'$, and $\psi+\Ge\psi'$, we get to first order in $\Ge$
that
\beq E\psi'=H\psi'+(V'-E')\psi, \eeq{p2}
and \eq{p1} implies
\beq \int \psi'(\Bx)\overline{\psi(\Bx)}+\psi(\Bx)\overline{\psi'(\Bx)}\, d\Bx=0, \eeq{p3}
i.e. that $\overline{\psi(\Bx)}$ is a linear combination of all the other eigenstates of $H$, as is well known. The restriction \eq{p3}
can also be viewed as a way of removing the nonuniqueness in the solution to \eq{p2} as otherwise we can add to $\psi$ any multiple of a solution
to the homogeneous equation, i.e., any multiple of $\psi$. Multiplying \eq{p2} by  $\overline{\psi(\Bx)}$ and using the fact that $H$ is Hermitian
gives 
\beq E'= \int \overline{\psi(\Bx)}\psi(\Bx)V'(\Bx)\, d\Bx. \eeq{p4}
Having determined $E'$ we can solve \eq{p2}, which is a Schr{\"o}dinger equation with the source term $(V'-E')\psi$. Equivalently, one can
rewrite it as
\beq \underbrace{\begin{pmatrix}\Bq' \\\Div\Bq'\end{pmatrix}}_{\BJ}=\BL\underbrace{\begin{pmatrix}\Grad \psi' \\ \psi'\end{pmatrix}}_{\BE}
-\underbrace{\bpm 0 \\ V'(\Bx)-E' \epm}_{\Bs},
\eeq{p5} 
where $\BL$ and $\BGG_1$ remain as before, but \eq{p3} must be satisfied to ensure uniqueness of the solution. Of course, a similar analysis
applies to all the other wave equations when there is no loss. 

One can once again make a transformation of the Cherkaev-Gibiansky type, to obtain a variational principle for the
time harmonic multielectron Schr{\"o}dinger equation with sources when the energy $E$ is complex, $E=E'+E''$. The simplifying analysis in Section
13.2 of \cite{Milton:2016:ETC} extended to the full complex wavefunction (not just its real part) implies that the functional
\beq W(\psi(\Bx)) =\int
[\Bp(\Bx)-\Bs(\Bx)]\overline{[\Bp(\Bx)-\Bs(\Bx)]}
+(E'')^2\,d\Bx,
\eeq{varsh}
where
\beq
\Bp(\Bx)=\Div\BA\Grad\psi(\Bx)+(E'-V(\Bx))\psi(\Bx),\quad \int\psi(\Bx)\overline{\psi(\Bx)}\,d\Bx=1,
\eeq{x18vp}
is minimized when $\psi(\Bx)$, constrained by its symmetries, satisfies the Schr{\"o}dinger equation with a (possibly complex) source term $\Bs$.
Here the overline denotes complex conjugation. Clearly we gain nothing when $E''=0$ and $E$ lies on the discrete spectrum, as the minimum occurs when $\Bp(\Bx)=0$
corresponding to the usual Schr{\"o}dinger equation with a real energy. However, when $E$ does not lie on the spectrum it
allows one to generalize density functional theory to states that are not ground states, although currently its application seems difficult:
see Chapter 11 of \cite{Milton:2016:ETC}.

\section{Desymmetrized form of the multielectron Schr{\"o}dinger equation}
\setcounter{equation}{0}
\labsect{11}

These equations are appropriate when 
the potential $V(\Bx)$ does not contain more than pairwise interactions between electrons. Due to the symmetries of the wavefunction
one can rewrite the time harmonic multielectron Schr{\"o}dinger equation in its equivalent desymmetrized form:
\beq  \underbrace{\bpm \Bp \\ \Gf \epm}_{\BJ}=\BL^D\underbrace{\bpm \Grad \psi \\ \psi \epm}_{\BE},\quad
\BL^D(\Bx_1,\Bx_2)\equiv\begin{pmatrix}-\BA & 0 \\ 0 & E-V^D(\Bx_1,\Bx_2)\end{pmatrix},\quad \BGG_1\BE=\BE,\quad \BGG_1\BJ=0,
\eeq{x18a}
where the superscript $D$ signifies desymmetrized,  $V^D(\Bx_1,\Bx_2)$ corresponds to the pair potential, not necessarily a function of just $\Bx_1-\Bx_2$,
$\BGG_1(\Bk)$ is the same as in \eq{x17}, and the wavefunction $\psi(\Bx)$ satisfies the usual symmetry properties.
The important feature is that $\BL^D$ only depends on the two electron coordinates $\Bx_1$ and $\Bx_2$ not the other electron coordinates.
To make the connection with the regular multielectron Schr{\"o}dinger equation, we introduce a symmetrization projection operator
\beq
\BGL=\bpm \BGL_A & 0 \\ 0 & \GL_a \epm,
\eeq{de1}
where $\GL_a$ acts on scalar fields and projects onto ones having the symmetries \eq{sy1} while
$\BGL_A$ acts on vector fields and projects onto ones having the symmetries \eq{sy2}. Then multiplying the desymmetrized Schr{\"o}dinger equation
\eq{x18a} on the left by $\BGL$ and comparing it with \eq{15.16} allows us make the identifications
\beq \BL=\BGL\BL^D\BGL, \quad \bpm \Bq \\ \Div\Bq \epm  =  \BGL\bpm \Bp \\ \Gf \epm.
\eeq{de2}
This last relation can be viewed as the differential constraints on $\BJ$ in \eq{x18a} that are implicit in the form of $\BGG_1(\Bk)$. With this
identification $\BL$ includes interactions between all pairs of electrons.
We do not require $\BJ$ to have any symmetry properties,
but we free to assume that $\BJ$ shares the same symmetries as $\BE$ with respect to interchange of
$\Bx_3$, $\Bx_4$,...$\Bx_N$ (but not with respect to $\Bx_1$ and $\Bx_2$).  This makes the action of $\BGL$ easier to compute.

For example, in a three electron system the action of $\GL_a$ on $\Gf$ is given by
\beqa
\GL_a\Gf (\Bx_1,\Bx_2,\Bx_3) & = &
\frac{1}{6}\Big[\Gf (\Bx_1,\Bx_2,\Bx_3)-\Gf (\Bx_2,\Bx_1,\Bx_3)+\Gf (\Bx_2,\Bx_3,\Bx_1)\nonumber\\
&&-\Gf (\Bx_3,\Bx_2,\Bx_1)+\Gf (\Bx_3,\Bx_1,\Bx_2)-\Gf (\Bx_1,\Bx_3,\Bx_2)\Big],
\eeqa{C6}
and  the action of $\BGL_A$ on $\Bp$ is given by
\beq \BGL_A\Bp=\bpm q_{1}(\Bx_1,\Bx_2,\Bx_3)\\ q_{2}(\Bx_1,\Bx_2,\Bx_3)\\ q_{3}(\Bx_1,\Bx_2,\Bx_3)\epm,
\eeq{CCA}
where
\beqa
q_{1}(\Bx_1,\Bx_2,\Bx_3)
& = & [p_{1}(\Bx_1,\Bx_2,\Bx_3)
- p_{1}(\Bx_1,\Bx_3,\Bx_2)- p_{2}(\Bx_2,\Bx_1,\Bx_3) + p_{2}(\Bx_3,\Bx_1,\Bx_2)- p_{3}(\Bx_3,\Bx_2,\Bx_1)+ p_{3}(\Bx_2,\Bx_3,\Bx_1)]/6, \nonum
q_{2}(\Bx_1,\Bx_2,\Bx_3)
& = & [p_{2}(\Bx_1,\Bx_2,\Bx_3)
- p_{2}(\Bx_3,\Bx_2,\Bx_1)- p_{1}(\Bx_2,\Bx_1,\Bx_3) + p_{1}(\Bx_2,\Bx_3,\Bx_1)- p_{3}(\Bx_1,\Bx_3,\Bx_2)+ p_{3}(\Bx_3,\Bx_1,\Bx_2)]/6, \nonum
q_{3}(\Bx_1,\Bx_2,\Bx_3)
& = & [p_{3}(\Bx_1,\Bx_2,\Bx_3)
- p_{3}(\Bx_2,\Bx_1,\Bx_3)- p_{1}(\Bx_3,\Bx_2,\Bx_1) + p_{1}(\Bx_3,\Bx_1,\Bx_2)- p_{2}(\Bx_1,\Bx_3,\Bx_2)+ p_{2}(\Bx_2,\Bx_3,\Bx_1)]/6, \nonum
&~&
\eeqa{C4}
with obvious extensions to when there are more than three electrons. More generally, assuming $\Gf $ has the expected
symmetries with respect to interchanges of $\Bx_3$, $\Bx_4$,...$\Bx_N$ then, as proved in Section 12.8 of  \cite{Milton:2016:ETC},
the action of $\GL_a$, is given by
\beqa &~&\GL_a\Gf (\Bx_1,\Bx_2,\Bx_3,\ldots,\Bx_N) = \nonum
&&\quad\frac{2}{N(N-1)}\Bigg[\Gf (\Bx_1,\Bx_2,\Bx_3,\ldots,\Bx_N)+
\sum_{i=3}^N(-1)^{i+1}\Gf (\Bx_2,\Bx_i,\Bx_1,\Bx_3,\Bx_4,\ldots,\Bx_{i-1},\Bx_{i+1},\ldots,\Bx_N)\nonum
&&-\sum_{i=3}^N(-1)^{i+1}\Gf (\Bx_1,\Bx_i,\Bx_2,\Bx_3,\Bx_4,\ldots,\Bx_{i-1},\Bx_{i+1},\ldots,\Bx_N)\nonumber\\
&&+\sum_{i=3}^N\sum_{\ell=i+1}^N(-1)^{i+\ell+1}\Gf \Big(\Bx_i,\Bx_\ell,\Bx_1,\Bx_2,\Bx_3,\ldots,\Bx_{i-1},\Bx_{i+1}\ldots,\Bx_{\ell-1},\Bx_{\ell+1},\ldots,\Bx_N\Big)\Bigg].
\eeqa{C5}
One does not need $\BGL_A$ to solve the equation iteratively by bouncing back and forth between real space and Fourier space, to satisfy respectively
the constitutive law and differential constraints respectively:
see Part  VI \cite{Milton:2020:UPLVI}
and Chapter 12 of \cite{Milton:2016:ETC}. The advantage
of the desymmetrized form of the  Schr{\"o}dinger equation is that going back to real space only requires the Fourier Transforms to be done on 
on the Fourier vectors $\Bk_1$ and $\Bk_2$, not on the Fourier vectors of the other electrons. This is because $\BL^D$ only depends on $\Bx_1$, $\Bx_2$,
and not on $\Bx_3$, $\Bx_4$,...$\Bx_N$.


\section*{Acknowledgements}
GWM thanks the National Science Foundation for support through grant DMS-1814854, and Christian Kern for helpful comments on the manuscript. The work was largely based on the
books \cite{Milton:2002:TOC, Milton:2016:ETC} and in the context of the latter book many thanks go to the friends and colleagues mentioned in
the acknowledgements of Part I \cite{Milton:2020:UPLI}. The author also thanks Vladimir Shalaev and Sergei Tretyakov for references
to early metamaterial related work. 

\ifx \bblindex \undefined \def \bblindex #1{} \fi\ifx \bbljournal \undefined
  \def \bbljournal #1{{\em #1}\index{#1@{\em #1}}} \fi\ifx \bblnumber
  \undefined \def \bblnumber #1{{\bf #1}} \fi\ifx \bblvolume \undefined \def
  \bblvolume #1{{\bf #1}} \fi\ifx \noopsort \undefined \def \noopsort #1{}
  \fi\ifx \bblindex \undefined \def \bblindex #1{} \fi\ifx \bbljournal
  \undefined \def \bbljournal #1{{\em #1}\index{#1@{\em #1}}} \fi\ifx
  \bblnumber \undefined \def \bblnumber #1{{\bf #1}} \fi\ifx \bblvolume
  \undefined \def \bblvolume #1{{\bf #1}} \fi\ifx \noopsort \undefined \def
  \noopsort #1{} \fi

\end{document}